\documentclass[physics,article,accept,pdftex,moreauthors]{Definitions/mdpi} 

\firstpage{1} 
\pubvolume{1}
\issuenum{1}
\articlenumber{0}
\pubyear{2023}
\copyrightyear{2023}
\datereceived{ } 
\daterevised{ } 
\dateaccepted{ } 
\datepublished{ } 
\hreflink{https://doi.org/} 

\def\hb{H$\beta$}

\def\rfe{R$_{\rm FeII}$}

\def\oiiib{[O{\sc iii}]$\lambda$5007}

\def\feii{Fe{\sc ii}}

\def\kms{km s$^{-1}$}
\def\cloudy{{\sc cloudy}}
\def\zsun{Z$_{\odot}$}

\usepackage{rotating}
\Title{Joint analysis of the iron emission \\ in the optical and near-infrared spectrum of I~Zw~1}

\TitleCitation{Joint analysis of the iron emission \\ in the optical and near-infrared spectrum of I~Zw~1}

\Author{Denimara Dias dos Santos $^{1,2,\ddagger *}$\orcidA{}, Swayamtrupta Panda $^{3,\ddagger *}$\orcidB{}, Alberto Rodríguez-Ardila $^{1,3}$\orcidC{}, Murilo Marinello $^{3}$\orcidD{}}

\date{Received } 

\AuthorNames{Denimara Dias dos Santos, Swayamtrupta Panda, Alberto Rodríguez-Ardila, Murilo Marinello}

\AuthorCitation{D. Dias dos Santos; S. Panda; A. Rodr\'iguez-Ardila; M. Marinello.}

\address{%
$^{1}$ \quad Divis\~ao de Astrof\'isica, Instituto Nacional de Pesquisas Espaciais (INPE), Avenida dos Astronautas 1758, S\~ao Jos\'e dos Campos, 12227-010, SP, Brazil.\\
$^{2}$ \quad Istituto Nazionale di Astrofisica (INAF), Osservatorio Astronomico di Padova, 35122 Padova, Italy\\ 
$^{3}$ \quad Laborat\'orio Nacional de Astrof\'isica (LNA), Rua dos Estados Unidos 154, Bairro das Na\c c\~oes. CEP 37504-364, Itajub\'a, MG, Brazil}

\corres{Correspondence: denimara.santos@inpe.br; spanda@lna.br}

\secondnote{These authors contributed equally to this work.}

\abstract{Constraining the physical conditions of the ionized media in the vicinity of an active supermassive black hole (SMBH) is crucial to understanding how these complex systems operate. Metal emission lines such as iron (Fe) are useful probes to trace the gaseous media's abundance, activity, and evolution in these accreting systems. Among these, the \feii{} emission has been the focus of many prior studies to investigate the energetics, kinematics, and composition of the broad-emission line region (BELR) from where these emission lines are produced.
In this work, we present the first simultaneous \feii{} modeling in the optical and near-infrared (NIR) regions. We use \cloudy{} photoionization code to simulate both spectral regions in the wavelength interval 4000 - 12000~\AA.
We compare our model predictions with the observed line flux ratios for I~Zw~1 - a prototypical strong \feii{}-emitting active galactic nuclei (AGN). This allows putting constraints on the BLR cloud density and metal content that is optimal for the production of the \feii{} emission, which can be extended to I~Zw~1-like sources, by examining a broad parameter space. We demonstrate the salient and distinct features of the \feii{} pseudo-continuum in the optical and NIR, giving special attention to the effect of micro-turbulence on the intensity of the \feii{} emission.}


\begin{document}




\section{Introduction}

Active Galactic Nuclei (AGNs) are harbored within the central cores of galaxies and host active supermassive black holes (SMBHs) at their very centers. The radiation originates from the heating of material that is drawn toward the SMBH and subsequently triggers ionization within the gas and metal-rich environment surrounding these systems \citep{Netzer_2015_AGN_Review, Padovani_etal_2017_AGN_Review}. The appearance of AGNs can vary based on the observer's perspective due to their intricate geometry \citep{Collin_etal_2006, panda2019quasar}. This phenomenon also becomes evident in their observed spectra.

AGNs are further classified in terms of the types of emission line profiles they demonstrate in their observed spectrum. Sources displaying a combination of broad (permitted and semi-permitted) as well as narrow forbidden emission line profiles are categorized as Type I sources, while those lacking broad emission lines are classified as Type-II \citep{antonucci93,urry_padovani_1995,sulentic2000phenomenology, Netzer_2015_AGN_Review}. This study exclusively concentrates on Type-I sources and their emissions. The origin of the broad emission lines can be traced back to a region located around 0.01 - 0.1 parsecs away from the central ionizing source. This measurement is derived through the widely recognized technique of reverberation mapping, which has been applied to a substantial sample of over 200 AGNs to date \citep{blandford1982reverberation, peterson_etal_2004_RM, bentz2013low, du2015supermassive, grier2017structure, panda2019current, Shen_SDSS_2023arXiv230501014S}. This region, known as the broad line region (BLR), contributes significantly to the emissions observed in a typical Type-I AGN spectrum \citep{1993ApJ...402..469P, 1994ApJ...425..622P, 1994ApJS...93...73R, 1995ApJS...97..285K, 1997ApJ...484...92V, berk2001composite, 2003A&A...407..461K, glikman2006near, 2006ApJ...641..689V, Marziani_etal_2010, bentz2013low, du2015supermassive, 2015A&A...576A..73P, Marziani2018AQuasars, 2019ApJ...874....8M, 2020MNRAS.492.3580W, 2020ApJ...899...73F, 2020A&A...642A..59R, 2020ApJ...905...51S, 2022ApJ...936...75L, 2022MNRAS.516.2671P, 2022ApJS..263...10L, 2023ApJ...944...29B, 2023MNRAS.523..545D}. Notably, among the salient BLR emission features is the \feii{} emission, encompassing a wavelength range from ultraviolet (UV) to near-infrared (NIR) \citep{wills1985broad, verner1999numerical, marinello2020panchromatic}.

The emission from \feii{} is remarkable in the Type-I AGN spectra, primarily due to the aggregation of an extensive array of transitions, above 344,000 in number. The accumulation of the lines mimics a continuum and is hence known as a pseudo-continuum in an AGN spectrum \citep{Phillips_1977,verner1999numerical, sigut2003predicted, sigut2004theoretical}. The large number of emissions stemming from the \feii{} ion play a pivotal role in characterizing the energy distribution within the BLR - approximately 25\% of the total emission originating in the BLR is attributed to the \feii{} lines \citep{wills1985broad}. Succeeding investigations have confirmed the utility of \feii{} emission as a surrogate for quantifying metal abundances in the BLR of AGNs across a broad range of redshifts, contributing to insights into the evolutionary trends of metallic elements within these galaxies \citep{hamann1992age, baldwin2004origin, panda2019quasar, martinez2021cafe, sarkar2021improved}.

Another crucial outcome from studies centered on \feii{} pertains to the development of emission templates specific to \feii{}. From a purely observational standpoint, the pioneering work of \citet{boroson1992emission} marked the inception of extracting \feii{} emission from a quintessential AGN \feii{} emitter, I~Zw~1, which itself has a long and rich history \citep{Phillips_1976, Phillips_1977, Oke_Lauer_1979, Joly_1981, Halpern_Oke_1987, Laor_1997, Negrete_etal_2012, Park_etal_2022}. The \feii{} template was meticulously constructed by isolating and excluding all emission lines except in the \feii{}. This template-driven approach continues to be employed across various studies, serving as the foundation for quantifying the optical intensity of \feii{}. \citet{boroson1992emission} played a pioneering role in comprehending \feii{} emission. In the same article, the authors established a robust association between the strength of \feii{} and various attributes of both the Broad Line Region (BLR) and the Narrow Line Region (NLR). Employing Principal Component Analysis (PCA) - a dimensionality reduction technique, they identified multiple correlations between observed spectroscopic parameters. Eigenvector1 (EV1) was the paramount correlation space among them. The EV1 from their study shows a strong correlation between the optical \feii{} emission within the spectral range of 4434 - 4684\AA, (centered at 4570~\AA), the peak intensity of [O{\sc iii}]$\lambda$5007 and the Full Width at Half Maximum (FWHM) of broad component of the \hb{} emission profile. This led to the well-established FWHM(\hb{}) vs. \rfe{} connection\footnote{ \rfe{} is usually the ratio of the integrated \feii{} emission within 4434 - 4684\AA\ to the broad \hb{} emission.}, that is the well-known optical plane of the quasar main sequence \citep{Sulentic_2000ApJ...536L...5S, shen2014diversity, Marziani_etal_2010, Marziani2018AQuasars, panda2019quasar, 2023A&A...669A..83D, 2023MNRAS.525.4474M}. Within this framework, \feii{} emission not only plays a crucial role in discerning the underlying factors driving EV1 \citep{panda2019quasar} but also serves as a link between line strength and the emitting gas physics for a diverse population of Type-1 AGNs along the EV1 plane.

Regarding \feii{} templates, \citet{vestergaard2001empirical} progressed in the UV domain by extending the template methodology into the ultraviolet spectrum using a high-quality HST/FOS spectrum for I~Zw~1. Subsequent efforts have since created several templates, effectively characterizing UV-optical \feii{} emission across large samples of AGNs \citep{tsuzuki2006fe, kovavcevic2015connections, dong2010prevalence, dong2011controls, kovavcevic2010analysis}. \citet{kovavcevic2010analysis} devised an innovative optical \feii{} template through a semi-empirical approach, combining the observed \feii{} emission with theoretical predictions based on allowed transitions between energy levels for the ion. By meticulously measuring individual \feii{} multiplet groups, they achieved a better overall agreement and were successful in getting an improved estimate for the \rfe{}. Reproducing \feii{} emission in sources similar to I~Zw~1 has been challenging, as empirical templates struggle to replicate specific features in observed spectra. In a recent development, \citet{park2022new} introduced a new template based on the Mrk 493 spectrum to improve the modeling of \feii{} emission in I~Zw~1-like sources.

Significant strides in predicting near-infrared (NIR) \feii{} spectra were accomplished by \citet{sigut1998lyalpha} and \citet{sigut2003predicted} through their utilization of the Ly$\alpha$ fluorescence excitation mechanism to elucidate \feii{} emission within the wavelength range of, 8500 - 9500~\AA. This Ly$\alpha$ process exclusively governs the stimulation of energy levels reaching up to 13.6~eV, giving rise to the emission lines observed within this spectral segment. This assertion was corroborated by observations conducted by \citet{rodriguez2002infrared}. A semi-empirical template for NIR \feii{} emissions was later developed by \citet{garcia2012near}, combining an observed spectrum of I~Zw~1 with theoretical models from prior works by \citet{sigut1998lyalpha, sigut2003predicted, sigut2004theoretical}. More recently, this semi-empirical template was successfully applied in the NIR Fe {\sc ii} modeling in several AGNs spectra by \citet{Marinello2016, marinello2020panchromatic}.

It is important to note here that the NIR spectrum exhibits \feii{} emission lines that are either isolated or semi-isolated, in contrast to the UV-optical region (see Figure~\ref{fig:spectra}). This feature provides a distinct advantage by enabling a more precise and reliable determination of the \feii{} emission line properties. The \feii{} lines around the 1-micron region at $\lambda$9997, $\lambda$10502, $\lambda$10863 and $\lambda$11127 are the most intense among the \feii{} lines \citep{Rudy2000The1, rodriguez2002infrared, riffel20060}.

Recent research \citep{Marinello2016} has highlighted the intrinsic correlation between \feii{} emission in the optical and near-infrared spectral regions. Due to the intricacies of \feii{} ion's behavior and the challenges associated with modeling its emission, researchers have been prompted to explore alternative approaches for investigating the \feii{} emitting gas \citep{martinez2015and, panda2020cafe}. Observational and photoionization modeling works have been indicating that less complex ions, such as Ca{\sc ii} triplet ($\lambda\lambda$8498\AA, 8542\AA, and 8662\AA, also known as CaT), and O{\sc i} lines at $\lambda$8446 and $\lambda$11287, can serve as proxies for studying the spatial distribution of \feii{} emitting gas \citep{Marinello2016, rodriguez2002infrared, martinez2015and, panda2020cafe, Panda_2021_CaFe}.

The \feii{} emission is dependent on the BLR temperature and gas density, and also on other parameters such as composition and Brownian motion within the BLR cloud – high temperature ($\sim$5000 - 10000~K), relatively high density (log n$_{\rm H}$ = 10 - 12 cm$^{-3}$), super-solar abundances and micro-turbulence of the order of $\sim$100 \kms{} have been shown to positively impact the production of the \feii{} emission, especially in the UV and optical \citep{baldwin2004origin,bruhweiler2008modeling, Panda2018ModelingPlane, panda2019quasar}. In a recent study  \cite{sab}, we assessed the NIR Fe{\sc ii} emission in I~Zw~1 under low column density regime (10$^{22}$ cm$^{-2}$). In this current work, we make a comprehensive analysis of the influence of micro-turbulence on NIR emission, over a broad range of cloud densities and metal content, in conjunction with the optical region.

We aim to model the \feii{} emission to understand the line formation and the nature of the physical conditions in AGN with strong \feii{} emission, exploring how changing metal content and micro-turbulence effectively increases the \feii{} optical and NIR emissions simultaneously. For this purpose, we investigate three cases of micro-turbulence velocities within the ionized media, i.e., 0, 10, and 100 \kms{} and their impact on the optical and NIR \feii{} emission (based on the previous study by \citet{Panda_2021_CaFe} which focused only on recovering the optical \feii{} emission). The inclusion of the micro-turbulence parameter has been shown to augment the \feii{} emission line intensity in the optical and UV regions, due to the increase in the overall random, Brownian motion of the ions within the cloud \citep{Panda2018ModelingPlane, Panda_2019_warm_corona, sarkar2021improved}.
This is the first study that tests the effect of micro-turbulence and metal content on the production of NIR \feii{} emission, along with the optical \feii{} emission.

The paper is organized as follows: Section~\ref{sec:2} presents the observational spectroscopic data in the optical and NIR for I~Zw~1. Section~\ref{sec:3} describes the extraction of the \feii{} emission in the optical and NIR regime and estimation of the corresponding \feii{} strengths from the observed spectra. In Section~\ref{sec:4}, we demonstrate the modeling setup using the photoionization code {\sc cloudy} \citep{Ferland2017TheCloudy}. Results are highlighted in Section~\ref{sec:5}, followed by a brief discussion in Section~\ref{sec:6}. We summarize our findings from this study in Section~\ref{sec:7}.

\section{I~Zw~1 the prototypical \feii{} emitter}\label{sec:2}

The I~Zwicky~1 (I~Zw~1) is a nearby (z=0.061) Narrow-line Seyfert 1 galaxy (NLS1), considered a prototypical \feii{} emitter, and widely studied in the literature \citep{wills1985broad, Rudy2000The1, vestergaard2001empirical, Veron-Cetty2004TheIZw, Huang2019ReverberationMass}. NLS1s, a subclass of classical Type-I AGNs, display a narrower-to-broader range of H$\beta$ line widths compared to classical Seyfert-I galaxies, which typically feature FWHM(H$\beta$) $<$ 2000 \kms{} and a flux ratio, \oiiib{}/H$\beta < $3 \citep{OSTERBROCK1985TheGalaxies}. 

The I~Zw~1 spectrum has a rich historical record of UV, optical, and NIR studies \citep{boroson1992emission, vestergaard2001empirical,kovavcevic2010analysis, garcia2012near}. Recent reverberation mapping estimated the BLR's distance of the line-emitting BLR from the central ionizing source, R$_{\rm BLR}$=37.2~light-days \citep{Huang2019ReverberationMass}, a crucial parameter in our model that allows to minimize the degeneracies in our models.

To perform the optical and NIR analysis, we use the reduced and available spectra of the I~Zw~1 from previous works \cite{rodriguez2002infrared, riffel20060, garcia2012near, Marinello2016}. The NIR spectrum was observed using the 3.2 m IRTF telescope (NASA Infrared Telescope Facility) at Mauna Kea, Hawaii-USA in 2000 (PI: Rodríguez-Ardila). It employed the SpeX spectrograph in cross dispersion mode (SXD) covering the wavelength interval 0.8 - 2.4 microns, and photometric bands zJHK, with a spectral resolution, R = 2000 corresponding to the 0.8” $\times$ 15” slit. \citet{rodriguez2002infrared} and \citet{riffel20060} presented in detail the reduced spectrum and related observational information. The optical counterpart region was obtained from the published work of \citet{rodriguez2002infrared}. The optical spectrum was obtained at the CASLEO Observatory (Complejo Astronómico el Leoncito — San Juán, Argentina), employing the REOSC spectrograph in long slit mode covering the range 3500 - 6800 \AA\ with resolution is 0.10 \AA~pixel$^{-1}$.

\section{Optical and NIR Fe{{\sc ii}} Templates}\label{sec:3}

\begin{sidewaysfigure*}
    \centering
    \includegraphics[width=\textwidth]{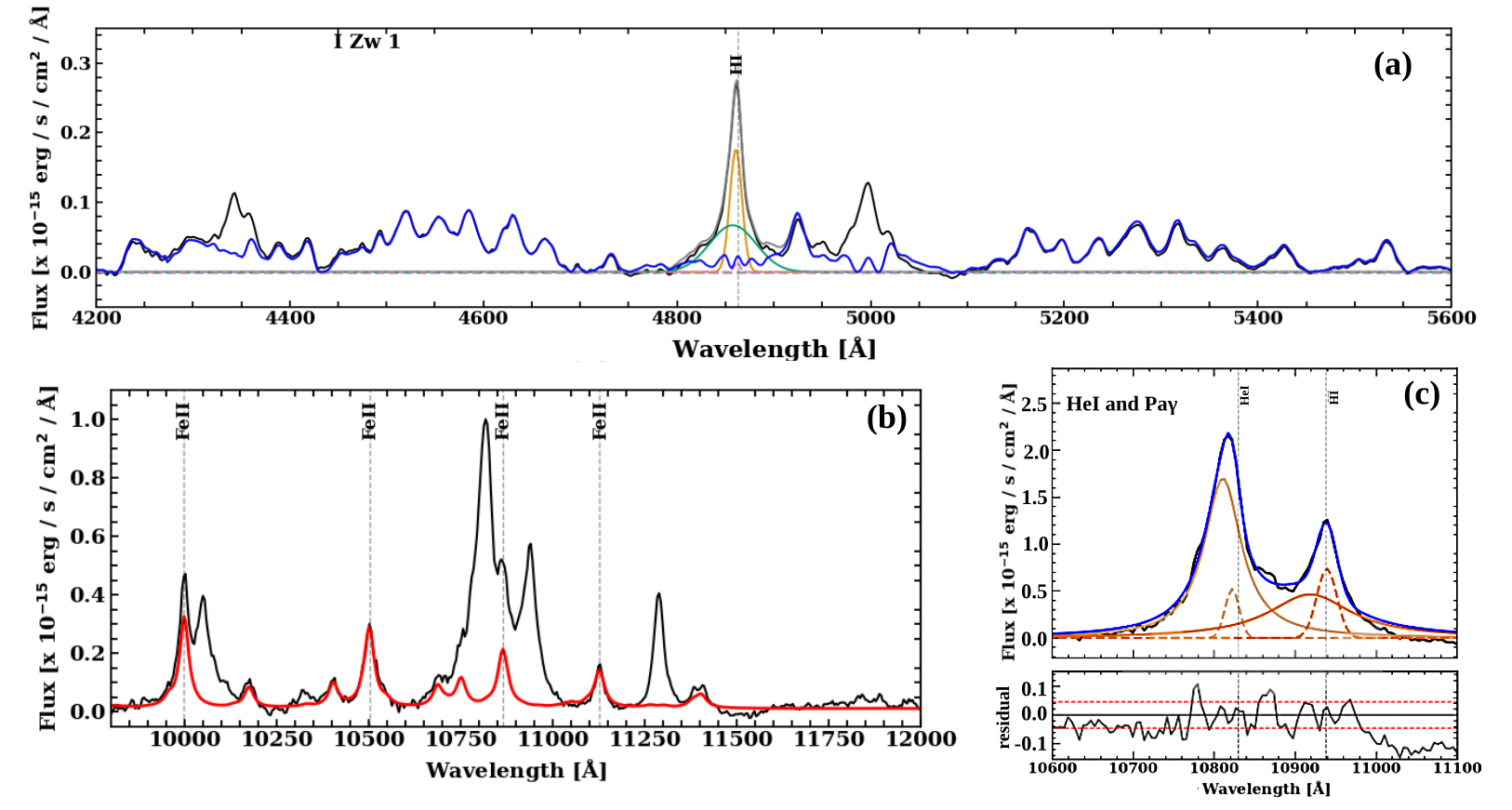}
    \caption{ Three panels depicting spectra of galaxy I~Zw~1. The (a) panel shows the optical spectrum ({the observed spectrum is shown in black}) with H$\beta$ emission line fits: green represents the outflow component, yellow is the BLR component, and orange-dashed indicates a faint narrow component. The blue curve represents the optical \feii{} template. The (b) displays the near-infrared spectrum in black with the fit of the semi-empirical \citet{garcia2012near} template in red, highlighting four main NIR \feii{} lines used to estimate the 1-micron intensity. In addition, (c) panel shows Pa$\gamma$-He{\sc i} line blend with He{\sc i}, where the solid line in orange represents the BLR components, {while the dashed lines imply} the NLR components.}
    \label{fig:spectra}
\end{sidewaysfigure*}

The \feii{} pseudo-continuum is comprised of a multitude of permitted Fe{{\sc ii}} lines spanning from the UV to the NIR spectral regions, necessitating the use of templates for accurate modeling. In the optical domain, we used the empirical Fe{{\sc ii}} template by \citet{boroson1992emission} derived from the spectrum of I~Zw~1, the archetypal Narrow-Line Seyfert 1 (NLS1) galaxy.
In the NIR range, we employed the semi-empirical template from \citet{garcia2012near}. These models encompass approximately 1915 lines, within the wavelength range of 8000 - 11600 \AA. The Fe{\sc ii} spectrum was synthesized using the template, where first we widened it based on the width of the 10502 \AA\ line, and then scaled the template to match with the observed spectrum. We use the 10502 \AA\ line as it is the most isolated \feii{} line in this region, thus allowing for accurate measurement of the overall \feii{} intensity. The best-fit template was determined by minimizing the chi-square value at, 10502 \AA. Figure~\ref{fig:spectra} illustrates the spectral fitting and component decomposition for the optical and NIR region for I~Zw~1. The estimation of flux uncertainty is based on the standard deviation of the best-fitted model, considering the spectrum within 3$\sigma$ of the lines of interest. For the ratios, we employed error propagation to estimate their uncertainties (see Table~\ref{tab1}). 
{In NLS1s, BLR {emission line} profiles are best represented using Lorentzian, while those of Type-I AGN with broader lines follow Gaussian shapes \citep{marziani2022intermediate}. We {fit} the emission lines using the {{\sc LMFIT}} library from Python, employing Lorentzian profiles to represent the BLR components, where the $\chi^2$ minimizes the residual left after subtraction of the {fitted} profile. For lines emitted by the Narrow-line Region (NLR), we employ only Gaussian components. This procedure enabled us to measure the integrated flux, FWHM, and centroid position of the lines {of interest for this study}. }


From these spectra and template {fits}, we obtain the optical \feii{} and NIR intensities {and their respective ratios to the nearest Hydrogen line}, where for the optical \feii{} we obtained R$_{\rm 4570}$ = 1.62$\pm$0.06. Our findings are similar to previous works ($\sim$1.47 \citep{sulentic2000phenomenology}). {For R$_{1\mu\,m}$, we observe a discrepancy between our findings and the results reported by \citet{Marinello2016} {where they recover a ratio,} R$_{1\mu m}$ = 1.81$\pm$0.08. It is crucial to clarify that the value of 1.81 is derived from the study by \citet{Marinello2016}, and not from our current investigation. {This value, from our analysis, is much smaller. We refer the readers to Table \ref{tab1}}. The noted disparity is attributed to differing methodologies employed in the two studies.} Specifically, \citet{Marinello2016} estimated measurements corresponding to Pa$\alpha$ and utilized the theoretical lines' ratio from \citet{garcia2012near} to re-scale to Pa$\beta$ intensity. In contrast, our approach involves the estimation of the flux of the broad component of Pa$\gamma$. {We opted for the utilization of Pa$\gamma$ rather than Pa$\beta$, as the latter is situated within a telluric spectral region \citep[see e.g.,][]{marinello2020panchromatic}}.

We investigated how the ratio between Pa$\gamma$ and Pa$\beta$ evolves under varying gas physical conditions from our \cloudy{} simulations (see Appendix~\ref{apendice}) instead of using the theoretical value for Pa$\beta$/Pa$\gamma$ (i.e., 0.8531). In Figure~\ref{apendice}, we highlight the dependence of the ratio on the cloud density as a function of the {a range of} metal content that is assumed in this work {using {\sc cloudy} photoionization code}. The value of the {Pa$\beta$/Pa$\gamma$} ratio varies between {1.05 - 3.38}. We assume a fixed column density (=10$^{24}$ cm$^{-2}$) with no micro-turbulence for this analysis. We then derive the R$_{1\mu m}$ estimates based on the calculated minimum and maximum value of the ratios obtained from {these} simulations {which we detail in the next section}. We re-scale the Pa$\gamma$ observed flux to get the flux for Pa$\beta$ (for the minimum and maximum values of their ratio) and estimate the minimum and maximum values for the intensity of \feii{} 1-micron lines. Consequently, we obtained a range of values for R$_{1\mu m}$ = 0.46$\pm$0.47, and 1.48$\pm$0.15, respectively. 


\begin{table}[]\scriptsize
\caption{Values measured from the near-infrared and optical spectra of I~Zw~1, where the fluxes are in units of {10$^{-14}$} erg~s$^{-1}$~cm$^{-2}$~\AA$^{-1}$. The errors for 1-micron values are respectively 0.15 and 0.47, derived from the values of 1.48 and 0.46. The flux errors are also reported side-by-side with their respective measured values. }
\label{tab1}
\begin{tabular}{lllllllll}
Fe{\sc ii}$\lambda$10502             & Fe{\sc ii}$\lambda$9998              & Fe{\sc ii}$\lambda$10863             & Fe{\sc ii}$\lambda$11127             & Pa$\gamma$                    & R$_{1\mu m}$             & H$\beta$                     & Fe{\sc ii} $\lambda$4570   & R$_{4570}$                \\ \hline\\
\multicolumn{1}{c}{2.70$\pm$1.38} & \multicolumn{1}{c}{4.13$\pm$1.06} & \multicolumn{1}{c}{0.150$\pm$0.17} & \multicolumn{1}{c}{2.32$\pm$1.29} & \multicolumn{1}{c}{5.96$\pm$0.83} & \multicolumn{1}{c}{\tiny{0.46$\pm$0.47 - 1.48$\pm$0.15}} & \multicolumn{1}{c}{10.30$\pm$1.84} & \multicolumn{1}{c}{16.70$\pm$2.80} & \multicolumn{1}{c}{1.62$\pm$0.06} \\ \hline
\end{tabular}
\end{table}

\section{Photoionization modelling} \label{sec:4}

The range of the physical conditions, used in this work, was based on previous approaches \citep{panda2020cafe,Panda_2021_CaFe, sab}. We employ the default model for the \feii{} ion in {\sc cloudy}, i.e., the one defined by \citet{verner1999numerical}, which includes 371 levels ranging in energy up to 11.6 eV, and 68,635 transitions to  {keep the consistency} with previous studies in low column density regime\cite{sab}.

We carry out {numerical} simulations using \cloudy{} v17.03 \citep{Ferland2017TheCloudy}. To constrain the number of free parameters to three, we adopt the following fixed values: the luminosity at 5100~\AA~, L$_{5100}$ = 3.19$\times$10$^{44}$ erg s$^{-1}$ from \citet{Kaspi2000ReverberationNuclei}, the BLR radius determined via reverberation mapping, R${_{\rm BLR}}$ = 37.2~light-days, based on \citet{Huang2019ReverberationMass}, and a cloud column density N$_{\rm H}$ = 10$^{24}$ cm$^{-2}$ {as concluded from earlier studies involving the recovery of the optical \feii{} emission in this source \citep{panda2020cafe, panda2021physical}}. Also, we adopt the spectral energy distribution for I~Zw~1 from  \citet{panda2020cafe} as the radiation field in {our} simulations. Thus, the {remaining free} parameters {are}: (i) the cloud mean hydrogen density, n$_{\rm H}$ {which ranges between} 10$^{9}$ - 10$^{14}$ cm$^{-3}$ and (ii) the metal content in the {cloud in the} 
range of 0.1 $\leq Z \leq$ 10 (in solar units), estimated using the \textit{GASS10} module \citep{Grevesse2010TheSun}. Moreover, we incorporate two scenarios for internal random motions within the BLR clouds, specifically micro-turbulence (V$_{turb}$) of 10 and 100 \kms{}.
{The micro-turbulence in the BLR is caused by random motions of photons within the ionized gas, which is likely triggered by magnetic fields in the confined clouds \citep{bottorf2000, ress1987} and acts as a secondary contributor to the overall line width, as has been shown in previous studies (e.g., \citet{kollatschny2013shape}). We assume the values for the micro-turbulent velocity based on previous works, which were successful in modeling the \feii{} emission in the UV region for I~Zw~1 about 10-30\kms{} \citep{bruhweiler2008modeling}. In addition, \citet{Panda2018ModelingPlane, panda2021physical} find that a micro-turbulence velocity between 10-100 \kms{} can reproduce the \feii{} emission in the optical region.} 
Additionally, as we aim to compare the effect of using micro-turbulence in the gas, we perform simulations in the absence of micro-turbulence in the gas (V$_{turb}$ = 0 \kms{}). This results in a total of 609 model combinations: V$_{turb}$ (3) $\times$ n$_{\rm H}$ (29) $\times$ Z (7).

\section{Results}\label{sec:5}

We derive from \cloudy{} models the optical \feii{} and the NIR \feii{} ratio, denoted as R$_{\rm 4570}$ and R$_{{1\mu}m}$, respectively. The R$_{\rm 4570}$ represents the ratio of the flux of optical \feii{} multiplets 37, 38, within the range 4434 - 4684~\AA\ to the flux of \hb{} broad component \citep{boroson1992emission, sulentic2000phenomenology}. The R$_{{1\mu}m}$ is calculated as the ratio of the combined fluxes from the four NIR prominent isolated \feii{} lines at wavelengths $\lambda$9997, $\lambda$10502, $\lambda$10863, and $\lambda$11127 to the flux of Pa$\beta$ broad component \citep{rodriguez2002infrared}. \citet{rodriguez2002infrared} and \citet{Marinello2016} demonstrate that the optical and NIR intensities are correlated, indicating that both emissions {should} have origin and excitation mechanisms in common.

\begin{sidewaysfigure}
\centering{\includegraphics[width=\columnwidth]{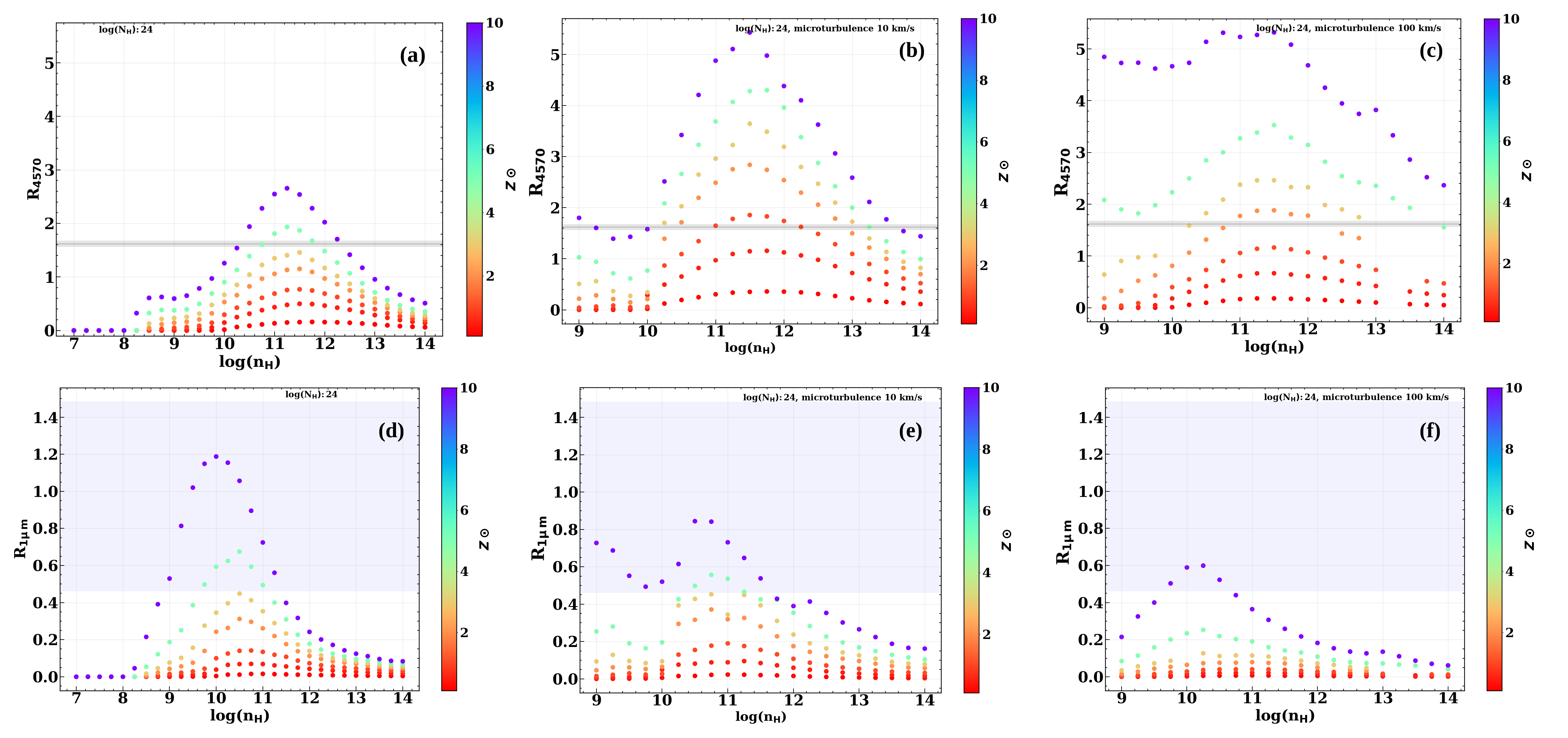}}
\caption{{Diagnostic diagrams from simulations depicting R$_{4570}$ intensities vs. n$_{\rm H}$ for a column density of 10$^{24}$ cm$^{-2}$. The upper panels (a, b, and c) illustrate different micro-turbulent velocities, with panels {(a) and (d) using 0 \kms{}, panels} (b) and (e) corresponding to 10 \kms{}, and panels (c) and (f) to 100 \kms{}. The color bars in all cases represent metallicities ranging from 0.1 M$_{\odot}$ to 10 M$_{\odot}$. The lower panels (d, e, and f) follow the same scheme but focus on R$_{{1\mu}m}$ along the y-axis. The dark gray line represents the observed value of R$_{4570}$, and the light gray bar symbolizes the associated error margin. In these lower panels, the blue shaded region denotes the estimated values for R$_{{1\mu}m}$ derived from the minimum and maximum Pa$\beta$/Pa$\gamma$ ratios obtained using {\sc cloudy} simulations.}}
\label{fig:diagram}
\end{sidewaysfigure}

In the optical \feii{} emission, the micro-turbulence effect has been extensively studied \citep{bruhweiler2008modeling, Panda_2021_CaFe}, but not in the NIR. In this work, for the first time, we {study} the micro-turbulence effect applied to the NIR \feii{} region. The results are shown in Figure~\ref{fig:diagram} for the column density scenario of 10$^{24}$ cm$^{-2}$ for optical {(upper panels)} and NIR {(lower panels)} spectral regions. The diagnostic diagrams in the same figure show the outcomes for V$_{turb}$ values of {0, 10, and 100 \kms{}}, for R$_{\rm 4570}$ and R$_{{1\mu}m}$ intensities as a function of {the local cloud} density. In each of the panels in Figure \ref{fig:diagram}, we observe how the intensities vary with different metal content levels in the BLR clouds {depicted using the color gradient}. {We further highlight the observed ratios, R$_{\rm 4570}$ and R$_{{1\mu}m}$, and their associated uncertainties, obtained by fitting the observed spectra, using the gray and blue shaded regions, respectively.}

\subsection{No micro-turbulence}
Our result without micro-turbulence (i.e., at 0 \kms{}) is shown in the left panels of Figure~\ref{fig:diagram}, where we compare the simulated results from \cloudy{} with the observed values for optical and NIR, R$_{\rm4570}$ = 1.62$\pm$0.06 and R$_{{1\mu}m}$ = 0.46–1.48, respectively, represented by the gray and blue bands\footnote{the symmetric errors for 1-micron values are 0.15 and 0.47, respectively{, also highlighted in Table \ref{tab1}}.}. Here, the widths of the gray bands account for the uncertainties associated with the ratios. The results show that we can reproduce R$_{\rm 4570}$ with a metal content above 3 Z$_{\odot}$ and with a local hydrogen density between 10$^{10.75}$–10$^{12.50}$ cm$^{-3}$. Moreover, from the NIR results, to reproduce R$_{{1\mu}m}$, it is necessary to have a local hydrogen densities range from 10$^{9.00}$ to 10$^{11.50}$ cm$^{-3}$, and a metal content above 3 Z$_{\odot}$. These results demonstrate a comparable physical parameter range to reproduce the {\feii{} strengths for} I~Zw~1 in the optical and NIR regimes within {our} setup.



\begin{figure*}
    \centering
    \includegraphics[width=\textwidth]{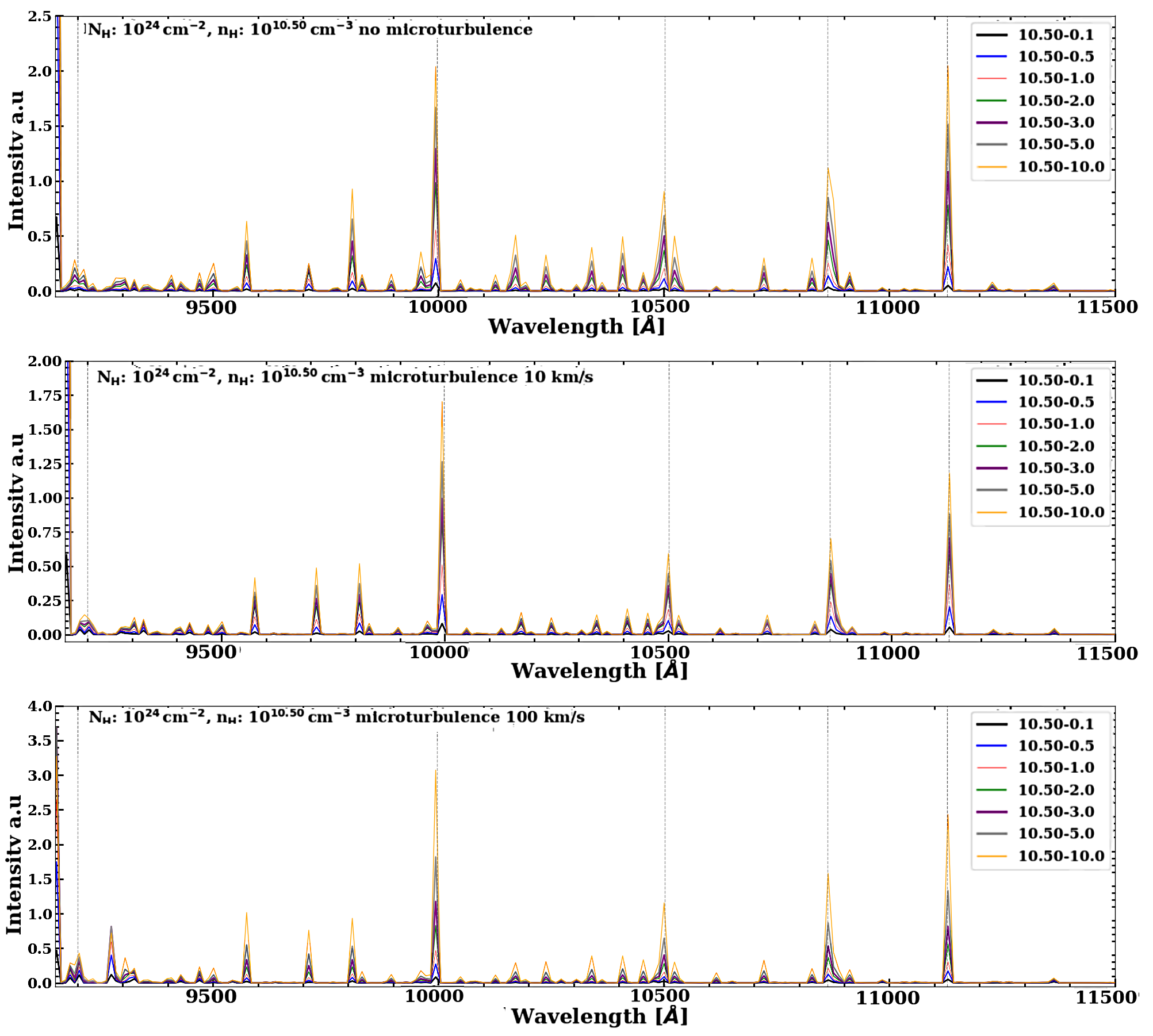}
    \caption{Predicted pseudo-continuum {from {\sc cloudy} simulations} for \feii{} emission for the representative case of n$_{\rm H}$=10$^{10.5}$ cm$^{-3}$ using 0.1, 0.5, 1, 2, 3, 5 and 10Z$_{\odot}$ for column density case 10$^{24}$ cm$^{-2}$. From top to bottom: zero micro-turbulence, 10, and 100 \kms{}, respectively.}
    \label{fig:comparation}
\end{figure*}

\subsection{Applying micro-turbulence}

Previous studies \citep{bruhweiler2008modeling,panda2020cafe} demonstrated that the value obtained {for R$_{\rm4570}$} in high metal content simulations can be achieved by increasing the micro-turbulent velocity. To investigate how changing metallicity and micro-turbulence can increase {both optical and} NIR emissions simultaneously, we include two cases of micro-turbulent velocity, at {10 and 100 \kms{}}, and compare it with the default case (i.e., with no micro-turbulence). We expect that the micro-turbulence in the gas {increases the Brownian motion within the cloud and allows for the \feii{} ions} to receive {preferentially more ionizing} photons, consequently leading to an increase in the intensity of the {\feii{} lines}. 

The results using micro-turbulence of {10 and 100 \kms{}} are shown in Figure~\ref{fig:diagram} in the middle and right panels, respectively. Our results for the optical region reveal that the \feii{} at 4570\AA\ strength increases {with the increase in} micro-turbulence, as expected from previous works \citep{panda2020cafe, Panda_2021_CaFe}.

On the other hand, the results in the NIR show a different behavior compared to the ones in the optical, as presented in Figure~\ref{fig:diagram}. 
Considering a single case with n$_{\rm H}$=10$^{12}$~cm$^{-3}$, solar metallicity, and V$_{turb}$ = 10 \kms{} (see Figure~\ref{fig:diagram} lower middle panel), R$_{{1\mu}m}$ is approximately $\sim$0.2, whereas at 100 \kms{}, it dropped to $\sim$0.05 (see Figure~\ref{fig:diagram} lower right panel). In contrast, the case with no micro-turbulence produces an R$_{{1\mu}m}$ value of $\sim$0.1 (see Figure~\ref{fig:diagram} lower left panel). Notably, the absence of micro-turbulence results in lower R$_{{1\mu}m}$ values compared to the scenario with V$_{turb}$ = 10\kms{}, as expected, based on the results obtained in the optical regime.
{Although}, for metallicities above solar, R$_{1\mu\,m}$ {values overall decrease} when compared to scenarios without micro-turbulence. {The R$_{1\mu\,m}$ appears to be suppressed and} fails to increase as expected from the optical results. For example, when we consider the highest metallicity value of 10\zsun{} (see Figure~\ref{fig:diagram}), and V$_{turb}$ = 10 \kms{}, and 100\kms{} the maximum \feii{} ratio in the NIR is, respectively, $\sim$0.8, and $\sim$0.6, whereas in the zero micro-turbulence case, this ratio approaches $\sim$1.2.

To comprehend this intriguing behavior from {our {\sc cloudy} simulations}, Figure~\ref{fig:comparation} provides a comparison of the \feii{} pseudo-continuum in the spectral region 9500–11500~\AA\ for a fixed local hydrogen density {(10$^{10.5}$ cm$^{-3}$) for} three scenarios: without micro-turbulence, V$_{turb}$ = 10 and 100 \kms{}. We note that the pseudo-continuum of V$_{turb}$ = 10 \kms{} exhibits less intense lines relative to the case without micro-turbulence. {The case with 100 \kms{} shows even weaker \feii{} lines.} {Although}, for the two cases with V$_{turb}$ = 10 and 100 \kms{}, we observe a boosting of lines {, specifically} between 9600–9900~\AA, which were not prominent in the spectrum in the zero micro-turbulence case.

\begin{figure}
\centering{\includegraphics[width=1\columnwidth]{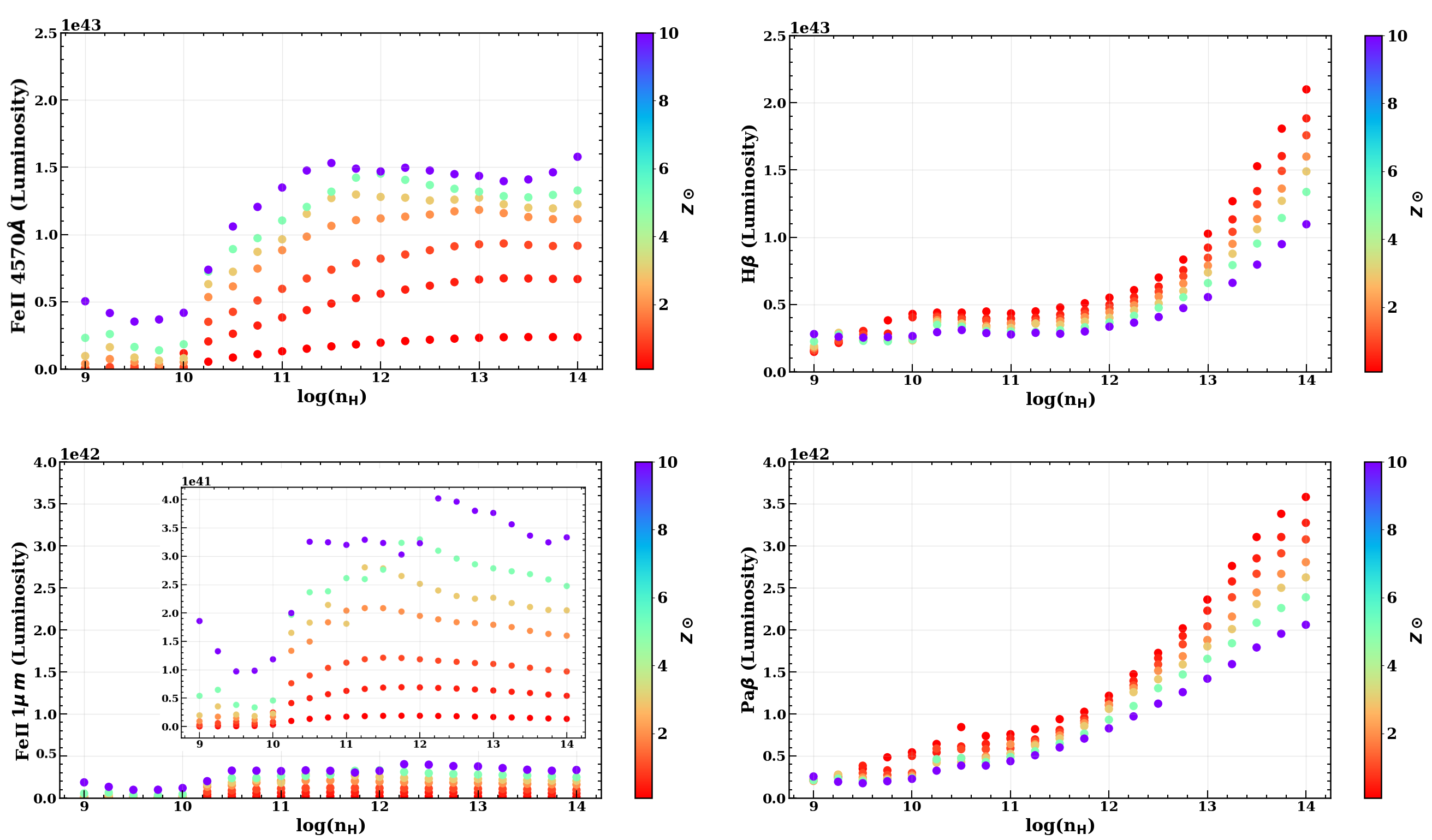}}
\caption{{Micro-turbulence case of 10 \kms{}. Top panel: The y-axis represented the luminosity of the 4570~\AA\ bump vs. local hydrogen density, and the luminosity of H$\beta$ line} vs. local hydrogen density. The colors corresponded to the metal content cases. Down panel: The colors and the x-axis represent the same meaning. The values on the y-axis differ from NIR quantities, respectively, the sum of the luminosity of the 4 \feii{} 1-micron lines, and Pa$\beta$ luminosity.}
\label{fig:hidrogen10}
\end{figure}
\begin{figure}
\centering{\includegraphics[width=1\columnwidth]{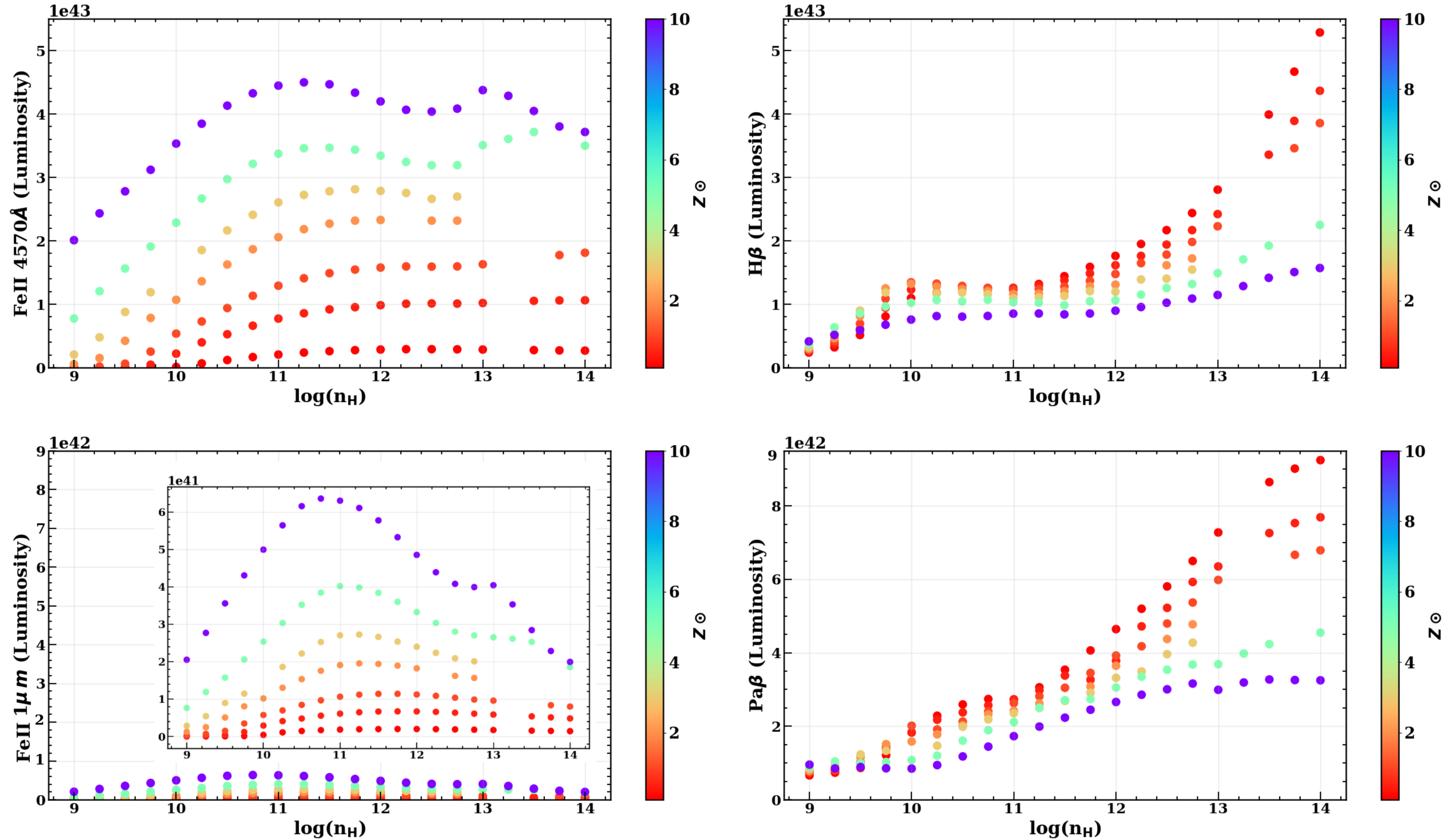}}
\caption{{Same from the previous figure caption, but for micro-turbulence case of 100 \kms{}.}}
\label{fig:hidrogen100}
\end{figure}

It is important to emphasize that the intensity of the NIR \feii{} is characterized by the ratio of four isolated emissions to Pa$\beta$ line, whereas the ratio in the optical is composed of blended \feii{} multiplets (m37, m38) normalized by the H$\beta$ \citep{kovavcevic2010analysis, marziani2021}. Hence, to understand the variations in the R$_{{1\mu}m}$, we need to study also the behavior of the Pa$\beta$ line across the parameter space, especially under the influence of microturbulence. We notice that the luminosity of the Pa$\beta$ emission significantly exceeds that of the NIR \feii{} lines. In contrast, in the optical range, both the H$\beta$ and the optical \feii{} lines exhibit a similar increase in luminosity with the density, as shown in Figures~\ref{fig:hidrogen10} and~\ref{fig:hidrogen100}.

{Moreover, the 1-micron lines exhibit two significant excitation mechanisms, collisional and Lyman-$\alpha$ fluorescence, whereas the \feii{} bump at 9200~\AA\ is exclusively influenced by the Lyman-$\alpha$ fluorescence mechanism \citep{rodriguez2002infrared, Marinello2016}. This distinction is an important indicator of the Lyman-$\alpha$ fluorescence mechanism. In our work, we note that the 9200~\AA\ region remains unaffected by the enhancement of micro-turbulence in the models.}



\section{Discussion}\label{sec:6}


The specific influence of micro-turbulence on the NIR emission of \feii{} is particularly intriguing in comparison to its well-documented impact in the optical region. The optical models suggest that when micro-turbulence is introduced in the gas, it enhances the intensity of the \feii{} lines. Surprisingly, our simulations indicated a contrary trend in the NIR region, where the R$_{{1\mu}m}$ displays lower values when microturbulence is considered than when micro-turbulence is absent.

The discrepancy between our observations and {{\sc cloudy}} predictions may raise intriguing questions regarding the true impact of micro-turbulence in the near-infrared {region}. Initially, it was anticipated that the increased micro-turbulence would result in enhanced photon absorption by \feii{} ions, leading to an increase in the emission in this spectral region as well, {as re-affirmed by the results from the \feii{} modeling in the optical region}. 
This {leads us to question} whether micro-turbulence potentially influences the Lyman-$\alpha$ absorption by the \feii{} ion, thereby affecting NIR \feii{} emission \citep{sigut1998lyalpha, rodriguez2002infrared, garcia2012near, Marinello2016}. A comprehensive investigation is required, which is beyond the scope of our current work.

We attribute the unexpected effects observed in the 1-micron lines {to a physical limitation,} where we {assume a fixed value for} the column density as a stopping criterion for the simulations. If we allow this parameter to be free, this can lead to a complex interplay between the column density and the metal content in the ionized gas, as found in \citet{Panda_2021_CaFe} in the case of the optical \feii{} emission. If this parameter is $\gtrsim$ $10^{24.5}$ cm$^{-2}$, one would have a cloud where the optical depth is $\gtrsim$ 1 {where} scattering effect becomes significant \citep[see][for more details]{Panda_2021_CaFe}. The limit at $10^{24}$ cm$^{-2}$ serves to constrain the cloud's physical condition, drawing inspiration from prior works \citep{Ferland_Persson_1989, bruhweiler2008modeling, Panda_2021_CaFe} wherein the radiative calculations are made under the optically thin regime. Thus, the non-monotonic behavior in the NIR may be ascribed to the collective influence of the microturbulence and cloud column density.{ In light of these intricate findings, further dedicated investigations are warranted to unravel the exact mechanisms at play, {that includes} also the advances in the {available \feii{} atomic datasets}.}

Coming back to the issue of the \feii{} flux ratio in the NIR obtained from the observed spectrum and under the physical conditions proposed in our study, our model is not able to replicate the ratio from \citet{Marinello2016}, which is {R$_{{1\mu}m}$} $\sim$1.81. To reproduce this value, a metal content exceeding 10\zsun{} would be required. This would mean that the optical and NIR \feii{} emissions require quite different metal compositions, even though they are formed in close vicinity to each other. This latter aspect is confirmed from the correlation observed for the FWHMs for \feii{} in the optical and NIR \citep{Marinello2016}. Nevertheless, exploring scenarios with significantly elevated metal content may be a possibility in future research of the NIR \feii{} emission, especially for high-accreting sources.

Furthermore, it is worthy to emphasize the importance of investigating how the hydrogen line ratios within the BLR of an AGN vary under different physical conditions. This inquiry will significantly improve the precision of measuring the \feii{} ratio in the NIR for quasars with a non-reliable Pa$\beta$ line (see \citet{Marinello2016, marinello2020panchromatic} for other examples). A future study should be considered, as it has the potential to further refine our understanding of AGN environments and improve the accuracy of critical astrophysical measurements.


\section{Conclusions}\label{sec:7}

In this study, we aimed at understanding the line formation and physical conditions in a strong \feii{} emitter AGN, I~Zw~1. We investigated how changes in micro-turbulence and metal content impact the optical and NIR emissions of \feii{} simultaneously. By introducing micro-turbulent velocities within the line-emitting cloud, of 10 and 100 \kms{}, employing a range of {cloud densities (10$^{9}$ - 10$^{14}$ cm$^{-3}$), and} metal content from sub-solar to 10 times solar inspired by previous works \citep{panda2019current, panda2020cafe}, we explored their combined effect on the recovery of the \feii{} intensities. Notably, this research is the first to examine the impact of micro-turbulence in the NIR regime, especially for the \feii{} emission, using photoionization model predictions and corroborating them with observed flux ratios obtained from archival spectra in optical and NIR for this source. While our findings align with prior findings made in the optical regime, we observed contrasting behavior in the NIR when micro-turbulence is introduced. Nonetheless, there is an overall agreement within the parameter ranges studied {in the two wavelength regimes}. A set of physical conditions can simultaneously reproduce optical and NIR \feii{} intensities, with n$_{\rm H}$ = 10$^{10.75}$ - 10$^{11.50}$ cm$^{-3}$, and a metal content ranging from 5 Z$_{\odot}$ $\lesssim$ Z $\lesssim$ 10 Z$_{\odot}$. 

{We reproduced the R$_{{1\mu}m}$ value for I~Zw~1 simultaneously in the optical and NIR. This result is significant not only for this source but is applicable also for I~Zw~1-like AGNs. The ability to replicate R$_{{1\mu}m}$ values across these wavelengths is a novel accomplishment. In conclusion, a comprehensive future analysis of excitation mechanisms and \feii{} models holds significant promise for further insights into this complex study.}

\vspace{6pt} 



\authorcontributions{The idea for the article and manuscript preparation was made by DDdS and SP. ARA and MM assisted with the editing and proofreading of the manuscript. All authors have read and agreed to the published version of the manuscript.}

\funding{This research received no external funding.}

\dataavailability{Data used in this work can be provided upon request to the corresponding authors.} 

\acknowledgments{The authors thank the Brazilian Agencies: Agency of Coordenação de Aperfeiçoamento de Pessoal de Nível Superior (CAPES), and Conselho Nacional de Desenvolvimento Cient\'{\i}fico e Tecnol\'ogico (CNPq).} 

\conflictsofinterest{The authors declare no conflict of interest.} 


\abbreviations{Abbreviations}{
The following abbreviations are used in this manuscript:\\

\noindent 
\begin{tabular}{@{}ll}
AGNs & Active galactic nuclei\\
BELR & Broad-emission line region\\
BLR & Broad-line region\\
EV1 & Eigenvector 1\\
FWHM & Full width at half maximum\\\
{FOS} & {Faint Object Spectrograph}\\
{HST} & {Hubble Space Telescope}\\
{IRTF} & {Infrared Telescope Facility}\\
NLR & Narrow-line region\\
NLS1 & Narrow-line Seyfert 1\\
NIR & Near-infrared\\
PCA & Principal component analysis\\
SMBHs & Supermassive black holes\\
SXD & Short wavelength cross-dispersed mode\\
UV & Ultraviolet\\
\end{tabular}
}

\appendixstart
\appendix

\section[\appendixname~\thesection]{}
\begin{figure}[H]
\centering{\includegraphics[width=1\columnwidth]{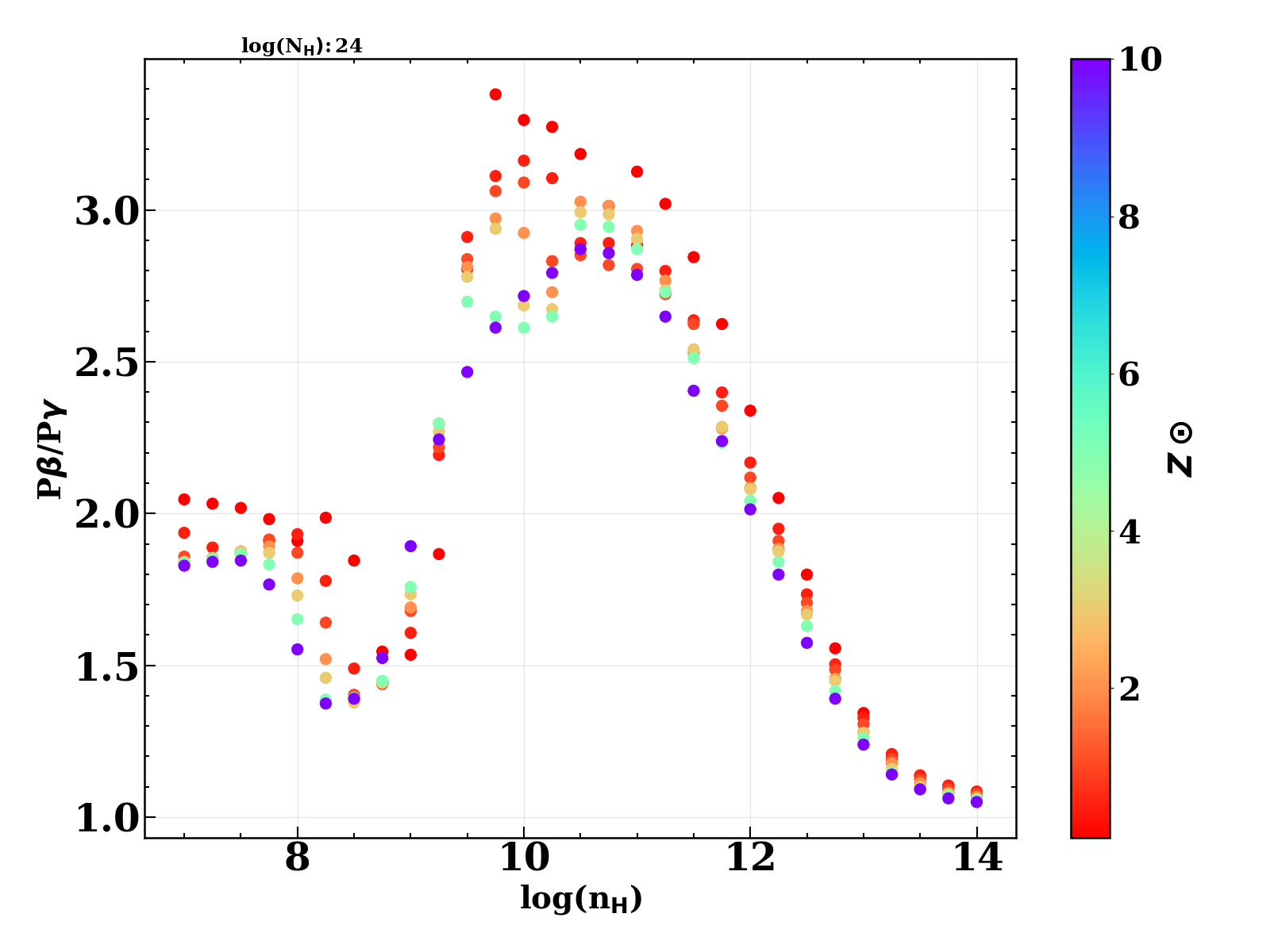}}
\caption{The y-axis represented the luminosity ratio for Pa$\beta$ to Pa$\gamma$ obtained from our simulations {for a range of} local hydrogen {densities} (x-axis). The colors correspond to the different metal content. The cloud column density is 10$^{24}$ cm$^{-2}$ at zero microturbulence.}
\label{apendice}
\end{figure}



\begin{adjustwidth}{-\extralength}{0cm}
\reftitle{References}
\bibliography{bib}

\PublishersNote{}
\end{adjustwidth}
\end{document}